\documentclass[prl,twocolumn,showpacs,floatfix,amsfonts,superscriptaddress]{revtex4}

\usepackage{graphicx,graphics,color,epsfig}
\usepackage{bm}
\usepackage{amsmath}
\usepackage{amssymb}

\newcommand{\beqa}{\begin{eqnarray}}
\newcommand{\eeqa}{\end{eqnarray}}

\usepackage{mathptm}    
\usepackage{dcolumn}                    
\usepackage{bm}                         
\usepackage{graphicx}

\usepackage{times}

\begin{document}

\title{Voltage dependence of Landau-Lifshitz-Gilbert damping of a spin in
a current driven tunnel junction}

\author{Hosho Katsura}
\email[Electronic address: ]{katsura@appi.t.u-tokyo.ac.jp}
\affiliation{Department of Applied Physics, University of Tokyo,
Hongo, Bunkyo-ku, Tokyo 113-8656, Japan}

\author{Alexander V. Balatsky}
\email{avb@lanl.gov}  \affiliation{Theoretical Division, Los Alamos
National Laboratory, Los Alamos, New Mexico 87545, USA}

\author{Zohar Nussinov}
\email{zohar@wuphys.wustl.edu}
\affiliation{Department of Physics, Washington University, St.
Louis, MO 63160, USA}

\author{Naoto Nagaosa}
\affiliation{Department of Applied Physics, University of Tokyo,
Hongo, Bunkyo-ku, Tokyo 113-8656, Japan} \affiliation{CERC, AIST
Tsukuba Central 4, Tsukuba 305-8562, Japan} \affiliation{CREST,
Japan Science and Technology Agency (JST)}

\begin{abstract}
We present a theory of Landau-Lifshitz-Gilbert damping $\alpha$ for
a localized  spin ${\vec S}$ in the junction coupled to the
conduction electrons in both leads under an applied volatege $V$. We
find the voltage dependence of the damping term reflecting the
energy dependence of the density of states. We find the  effect is
linear in the voltage and cotrolled by particle-hole asymmetry of
the leads.
\end{abstract}


\date{\today}
\maketitle

\section{Introduction}

Spintronics is an emerging subfield that holds the potential to
replace conventional electronic devices with spintronic analogues
where the manipulation, control, and readout of spins will
enable novel functionality with no or little electronic charge
dynamics \cite{spintronics}. In order to realize this
promise, the spin dynamics of the small scale devices needs to be well
controlled. One of the most pressing questions concerns
a set up which would preserve coherence and allow a manipulation of spins.
In most systems, the relevant spin degrees of freedom are coupled to
some bath, such as a fermionic bath of electrons. The detailed
dynamics of single spins when in contact with such a bath
plays a pivotal role in addressing decoherence in potential
spintronic systems.

\begin{figure}
\centerline{\includegraphics[width=0.9\columnwidth]{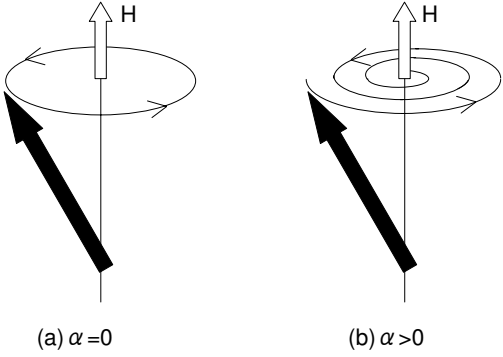}}
\caption{Sketch of the dissipative spin dynamics.
Panel (a) depicts a cartoon of the Larmor
precession of the spin about the direction of an applied magnetic
field (B). In panel (b), a caricature of the spin dynamics in the
presence of Landau-Lifshitz-Gilbert damping is shown.}
\label{FIG:FIG2}
\end{figure}

The conventional way to treat this problem is via a Caldeira-Leggett
approach where the external bath is modeled by collective
excitations which are capable of destroying coherent spin dynamics.
Often, spin dynamics is described by a Landau-Lifshitz-Gilbert equation
\cite{ll1,ll2}:
\begin{equation} {\frac {d{\vec S(t)}}{dt}} = - {\vec
S(t)}\times{\vec h} -\alpha {\vec S(t)}\times{\frac{d{\vec
S(t)}}{dt}}, \label{LLG2}
\end{equation}
where ${\vec h}$ is, up to constant prefactors, the external
magnetic field and the coefficient $\alpha$ captures the damping due
to the external bath. A caricature of the solution of this equation
\cite{Chika} is provided in Fig.\ref{FIG:FIG2}. There are standard
methods to calculate $\alpha$ in an equilibrium situation when, say,
one considers a spin in a Fermi liquid \cite{Heinrich,Halperin}.

In the current publication, we address a related novel question
concerning the effect of an applied voltage bias on the Gilbert
coefficient $\alpha$. Our work complements the recent results of
\cite{Onoda} wherein the effects of the ``retarded'' electronic
contributions in the equations of motion for a system of spins were
studied. Both such retarded correlations \cite{Zhu03} as well as
additional ``Keldysh'' correlations generally manifest themselves in
the single spin equations of motion, see e.g. \cite{nsabz} for
general spin equations of motion entailing the effects of both
correlations. In the current work, we examine the voltage dependence
of Gilbert damping. For the sake of clarity, we depart from the
Keldysh contour formalism of \cite{Onoda,Zhu03,nsabz}, and use
a Caldeira-Leggett approach.

\begin{figure}
\centerline{\includegraphics[width=0.9\columnwidth]{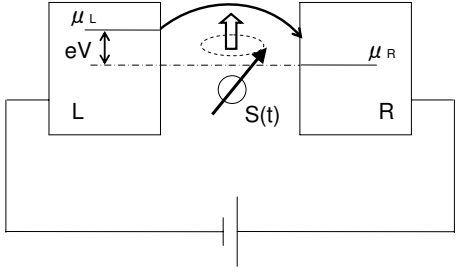}}
\caption{Magnetic impurity coupled to two electrodes. $\mu_{\rm L}$
and $\mu_{\rm R}$ denotes the chemical potentials of the left and
right leads respectively.  The voltage drop across the two
electrodes is $eV= \mu_{L} - \mu_{R}$.} \label{FIG:FIG1}
\end{figure}

In what follows, we consider the case of a junction between two
electrodes that contains one spin $\vec{S}$, see Fig.\ref{FIG:FIG1}.
This spin $\vec{S}$ may be the spin of a single magnetic impurity or
it may portray the spin of a cluster at low temperature when the
spins in the cluster are locked. Upon applying a finite bias between
the electrodes of Fig.\ref{FIG:FIG2}, a current flow is generated.
Thereafter, at long times, the system is at a steady but
non-equilibrium state so long as the voltage bias $V$ is applied. We
will focus on the voltage dependence of the damping term $\alpha(V)$
in Eq.(1). We find that the change in the density of states
associated with the chemical potential gradient across the junction
triggers a modification to the damping $\alpha$ that is {\em linear}
in voltage and is proportional to the particle-hole assymmetry of
the density of states. The scale of the correction is set by the
Fermi energy of the metal in the leads $E_F$ and by particle-hole
asymmetry in the density of states:
 \beqa \alpha(V) = \alpha_0+\alpha_1(V)
= \alpha_0 ( 1+ O(1)eV/E_F). \label{voltage+} \eeqa
 This result
vividly illustrates the presence of voltage induced damping in such
junctions. Spin unpolarized electrons tunneling across the junction
interact via exchange interaction with the spin $\vec{S}$ and
produce random magnetic fields that disorder the local spin. This
noise augments that already present equilibrium magnetic noise in a
Fermi liquid bath. Such a behavior of $\alpha(V)$ with the external
voltage is in line with the works of \cite{hooley}. An analysis of
a related single spin problem in a Josephson junction (instead of
the normal junction studied here) was advanced in
\cite{Zhu03,nsabz,bulaevskii}.

We will shortly derive the effective single spin action from which
the principle equation of motion of Eq.(\ref{LLG2}) follows. Several
technical details of our derivation are given in the appendices.


\section{THE SYSTEM AND OUR PRINCIPLE RESULTS}

The physical system under consideration in this publication is
illustrated in Fig.\ref{FIG:FIG1}. It consists of two (left(L) and
right(R)) electrodes across which a voltage bias is applied; a
magnetic impurity ($\vec{S}$) is situated in between (or lies on one
of) the electrodes. An external magnetic field ${\vec{B}}$ is
present. In the absence of effects stemming from conduction
electrons in the tunneling barrier, the single spin would precess at
the Larmor precession frequency about the applied field direction
(see, e.g. panel (a) of Fig.\ref{FIG:FIG2}). With the external
circuit elements present, the spin motion becomes dissipative (as
schematically shown in panel (b) of Fig.\ref{FIG:FIG2}).

With the spin embedded in the tunneling barrier, the work function
is modified and the conduction electron tunneling matrix element is
supplanted by a Kondo like exchange term $ J(\vec{S} \cdot
\vec{\sigma}_{c})$, with $\vec{\sigma}_{c}$ denoting the conduction
electron spin. In what follows, we will dispense with the $c$
subscript. The Hamiltonian governing this system is given by
\begin{eqnarray}
{\cal H} &=& {\cal H}_e + {\cal H}_s + {\cal H}_T,
\\
{\cal H}_e &=& \sum_{\alpha,k,\sigma} \xi_{\alpha k} c_{\alpha k
\sigma}^{\dagger}c_{\alpha k \sigma}, \nonumber
\\
{\cal H}_s &=& - {\vec h} \cdot {\vec S(t)}, \nonumber
\\
{\cal H}_T &=& {\frac{1}{\Omega}}{\sum_{\alpha,k,\sigma}}{\sum_{\beta,p,\sigma'}}
c_{\alpha k \sigma}^{\dagger} (T_{\alpha\beta})_{\sigma \sigma'}
c_{\beta p \sigma'}, \label{Ham}
\end{eqnarray}
where $c_{\alpha k \sigma}^{\dagger}~(c_{\alpha k \sigma})$ creates
(annihilates) an electron with momentum $k$ and spin $\sigma \in \{
\uparrow,\downarrow \}$ in the lead $\alpha \in \{{\rm L}, {\rm
R}\}$. The abbreviation $\xi_{\alpha k} = \epsilon_{\alpha k}
-\mu_{\alpha}$, where $\epsilon_{\alpha k}$ is the energy of
electron with momentum $k$ in (the lead) $\alpha$ and $\mu_{\alpha}$
is the chemical potential in (the lead) $\alpha$. The second term in
Eq.(\ref{Ham}), ${\cal H}_s$, is Zeeman energy of the spin in an
external magnetic field $\vec B$. Here, $\vec h \equiv g \mu_B {\vec
B}$ with $g$ gyromagnetic ratio and $\mu_B$ the Bohr magneton. The
last term in Eq.(\ref{Ham}), ${\cal H}_T$, represents both Kondo
coupling and direct tunneling process, where the amplitudes
$\{T_{\alpha \beta}\}=\{ T_{{\rm L}{\rm L}},T_{{\rm L}{\rm
R}},T_{{\rm R}{\rm L}},T_{{\rm R}{\rm R}} \}$ are tunneling matrix
elements and their explicit forms are
\begin{eqnarray}
T_{LL} &=& J_{LL}({\vec \sigma} \cdot {\vec S(t)}), \nonumber
\\
T_{RR} &=& J_{RR}({\vec \sigma} \cdot {\vec S(t)}), \nonumber
\\
T_{LR} &=& T_{RL} = (T_0 \delta_{\sigma \sigma'} + J_{LR}({\vec
\sigma} \cdot {\vec S(t)})). \label{tunnel_amp}
\end{eqnarray}
Here, $T_0$ is the direct tunneling matrix element and $J_{\alpha
\beta}$ is the Kondo coupling, 'while $\Omega$ denotes the Volume of
each lead (assumed, for simplicity, to be the same)'. Typically, from
the expansion of the work function for tunneling,
${J_{LR}}/{T_0} \sim J/U$, where $U$ is the height of a
spin-independent tunneling barrier and $J$ the magnitude of the spin
exchange interaction ~\cite{Zhu_Balatsky}. Typical values of the
ratio the spin dependent to spin independent tunneling amplitudes in
Eq.(\ref{tunnel_amp}), $J_{\alpha \beta}/T_{0}$, are
${\cal{O}}(10^{-1})$, with a typical Fermi energy $E_{F}^{R}$ of the
order of several electron-volts. From the Hermiticity of the
Hamiltonian, we can  find that the matrix element $(T_{\alpha
\beta})_{\sigma \sigma'}$ satisfies $((T_{\alpha \beta})_{\sigma
\sigma'})* = (T_{\beta \alpha})_{\sigma' \sigma}.$

In the up and coming, we derive the effective action for the single
impurity spin via an imaginary time path integral formalism. The
full action is given by
\begin{eqnarray}
S = \int_0^{\beta}d{\tau} \sum_{\alpha k \sigma} c_{\alpha k
\sigma}^{\dagger} \partial_{\tau} c_{\alpha k \sigma} +
iS\omega(\vec S(\tau)) +\int_0^{\beta}d{\tau}{\cal H}(\tau),
\label{action}
\end{eqnarray}
where the second, Wess-Zumino-Novikov-Witten(WZNW), term in Eq.(\ref
{action}) depicts the Berry phase accumulated by the spin
\cite{Fradkin}. In our action, we have the following quadratic form
of fermions,
\begin{eqnarray}
&& \int_0^{\beta} d{\tau} {\frac{1}{\Omega}} \sum_{\alpha k \sigma}
\sum_{\beta p \sigma'}  c_{\alpha k \sigma}^{\dagger}
(\delta_{\alpha \beta} \delta_{\sigma \sigma'}(\partial_{\tau} +
\xi_{\alpha k}) + (T_{\alpha \beta})_{\sigma \sigma'}) c_{\beta p
\sigma'} \nonumber
\\
&& \equiv \int_0^{\beta} d{\tau} {\frac{1}{\Omega}} \sum_{\alpha k
\sigma} \sum_{\beta p \sigma'} c_{\alpha k \sigma}^{\dagger}
((M_{\alpha \beta}(\tau))_{\sigma \sigma'})_{kp} ~ c_{\beta p \sigma'}.
\label{quad}
\end{eqnarray}
We may integrate over the lead electrons to obtain the effective
action for the spin
\begin{equation}
S_{\rm {eff}}(\vec S(\tau)) \sim i S \omega({\vec S(\tau)}) +
\int_0^{\beta} d{\tau} {\vec h} \cdot {\vec S(\tau)} -{\rm{ln}}{\rm {det}} M , \label{eff}
\end{equation}
where ${\rm {det}} M$ means \textit{functional determinant} of $M.$
From the third term in Eq.(\ref {eff}), we obtain a quadratic
non-local in time interaction of the spin with itself, $\vec
S(\tau)$ as
\begin{equation}
\Delta S(\vec S(\tau)) = -2 \int d{\tau} \int d{\tau'} {\vec S(\tau)}
\cdot {\vec S(\tau')} K(\tau-\tau'), \label{K+}
\end{equation}
where
\begin{eqnarray}
K(\tau) &=& \sum_{\alpha, \beta \in \{ \rm{L,R} \}} K_{\alpha
\beta}(\tau), \nonumber
\\
K_{\alpha \beta} &=& \sum_{k \in \alpha} \sum_{p \in \beta}
{\frac{J_{\alpha \beta} J_{\beta \alpha}}{2}} {\frac{1}{\beta}}
\sum_{\omega_m} {\frac {f(\xi_k)-f(\xi_p)}{i\omega_m +\xi_k -\xi_p}}
e^{-i \omega_m \tau}, \label{Kt}
\end{eqnarray}
with $f(\xi)$ denotes the Fermi distribution function (see APPENDIX
A). The effective action $\Delta S$ of Eq.(\ref{eff}) can be
decomposed into two (trivial and non-trivial) components as $\Delta
S = \Delta S_{\rm loc} + \Delta S_{\rm{dis}}$, with
\begin{eqnarray}
\Delta S_{\rm loc} &=& -2K(\omega=0) \int d{\tau} ({\vec S(\tau)})^2,
\nonumber
\\
\Delta S_{\rm{dis}} &=& \int d{\tau} \int d{\tau}' ({\vec
S(\tau)}-{\vec S({\tau})})^2 K(\tau-{\tau}'). \label{locdis}
\end{eqnarray}
Here, $K(\omega=0)$ is the zero-frequency Fourier component of
$K(\tau)$. The first term ($\Delta S_{\rm loc}$) is a trivial
constant as $S(\tau)^2 = S^2$. The nonlocal part ($\Delta
S_{\rm{dis}}$) represents the dissipative effect due to the coupling
between $S(\tau)$ and electrons bath. The integral kernel $K(\tau)$
is calculated in the same way as the Caldeira-Leggett theory
\cite{Leggett,Leggett2} leading to
\begin{equation}
K(\tau) = \int_0^{\infty} d{\omega} J(\omega) \frac{{\rm
cosh}({{\beta}/2}-|\tau|){\omega}}{{\rm sinh}({\beta \omega}/2)},
\label{K}
\end{equation}
where $J(\omega)$ is the \textit{spectral density} and its explicit
form is
\begin{equation}
J(\omega) = \sum_{\alpha \beta} \sum_{k \in \alpha} \sum_{p \in
\beta} {\frac{J_{\alpha \beta} J_{\beta \alpha}}{2}} [f(\xi_k)
-f(\xi_p)] \delta (\omega+ \xi_k -\xi_p). \label{spec}
\end{equation}
The details of the derivation of Eq.(\ref{spec}) are provided in
APPENDIX B.
The spectral density of Eq.(\ref{spec}), $J(\omega)$, is estimated
as
\begin{eqnarray}
J(\omega) &=& \sum_{\alpha \beta} {\frac{J_{\alpha \beta} J_{\beta
\alpha}}{2}} \int_{E_F^{\alpha}}^{\infty} d{\xi_{\alpha}}
N(\xi_{\alpha})\int_{E_F^{\beta}}^{\infty} d{\xi_{\beta}} N(\xi_{\beta})
\nonumber
\\
&& \times [f(\xi_{\alpha})-f(\xi_{\beta})]
\delta(\omega-\xi_{\alpha}-\xi_{\beta}) \nonumber
\\
&\sim& \sum_{\alpha \beta}{\frac{J_{\alpha \beta} J_{\beta
\alpha}}{2}} N(\xi_{\alpha}=0) N(\xi_{\beta}=0) \omega,
\label{spec2}
\end{eqnarray}
where $E_F^{\alpha}$ denotes the Fermi Energy of the lead $\alpha$.
It is obvious that $J(\omega)$ in Eq.(\ref{spec2}) is proportional
to $\omega$, i.e., $J(\omega)$ is Ohmic. If spectral density is
expressed as $J(\omega)={\eta \omega}/{2 \pi}$, then
the Gilbert coefficient $\alpha$ in Eq.(\ref{LLG2}) is exactly equal
to $\eta$. By varying the total action with respect to the spin
${\delta S}/{\delta {\vec S(\tau)}}=0$, we immediately obtain the
Landau-Lifshitz-Gilbert equation with $\alpha={\frac{2
\pi}{\omega}}J(\omega)$(see APPENDIX C). In other words, the voltage
dependence of $\alpha$ in Eq.(\ref{LLG2}) is identically the same as
that of $J(\omega)$. We next examine the voltage dependence of
$J(\omega)$.

If we apply a voltage leading to a chemical drop of
$\Delta\mu_L-\Delta\mu_R=eV$. Assuming, for example,that
 the
net charge on both right and left leads is unchanged, we also have
 $D^L(E_F)\Delta\mu_L+D^R(E_F)\Delta\mu_R=0$.
With these constraints we get
\begin{eqnarray}
\Delta\mu_L&=&\frac{D^R(E_F)}{D^L(E_F)+D^R(E_F)}eV,
\nonumber \\
\Delta\mu_R&=&-\frac{D^L(E_F)}{D^L(E_F)+D^R(E_F)}eV,
\end{eqnarray}
the Gilbert coefficient $\alpha$ may be approximated as\\
\begin{eqnarray}
\alpha(V) &\sim& 2\pi\Bigl(\frac{J_{LL}^2}{2} D^L(E_F+\Delta\mu_L)^2 + \frac{J_{RR}^2}{2}
D^R(E_F+\Delta\mu_R)^2 \nonumber\\
&&+J_{LR}^2 D^L(E_F+\Delta\mu_L) D^R(E_F+\Delta\mu_R)\Bigr)
\nonumber
\end{eqnarray}
\begin{eqnarray}
&\sim& 2\pi \Bigl(
\frac{[J_{LL}D^L(E_F)]^2}{2} + \frac{[J_{RR} D^R(E_F)]^{2}}{2} + J^2_{LR}D^L(E_F)D^R(E_F)\nonumber \\
&&+(J^2_{LL}D^L(E_F)\frac{\partial{D^L(E_F)}}{\partial{E_F}}
+   J^2_{LR}D^R(E_F)\frac{\partial{D^L(E_F)}}{\partial{E_F})})\Delta\mu_L
\nonumber \\
&&+(J^2_{RR}D^R(E_F)\frac{\partial{D^R(E_F)}}{\partial{E_F}}
+   J^2_{LR}D^L(E_F)\frac{\partial{D^R(E_F)}}{\partial{E_F})})\Delta\mu_R)
\Bigr) \nonumber
\end{eqnarray}
\begin{eqnarray}
&=&2\pi \Bigl(
\frac{[J_{LL}D^L(E_F)]^2}{2} + \frac{[J_{RR}D^R(E_F)]^{2}}{2}
+ J^2_{LR}D^L(E_F)D^R(E_F)\nonumber \\
&+& \frac{eV}{D^L(E_F)+D^R(E_F)}
(J^2_{LL}D^L(E_F)D^R(E_F)\frac{\partial{D^L(E_F)}}{\partial{E_F}}
\nonumber \\
&-& J^2_{RR}D^L(E_F)D^R(E_F)\frac{\partial{D^R(E_F)}}{\partial{E_F}}
\nonumber \\
&+& J^2_{LR}D^R(E_F)D^R(E_F)\frac{\partial{D^L(E_F)}}{\partial{E_F}}
\nonumber \\
&-& J^2_{LR}D^L(E_F)D^L(E_F)\frac{\partial{D^R(E_F)}}{\partial{E_F}})\Bigr).
\label{Gil}
\end{eqnarray}
The change in the density of states associated with the chemical
potential gradient across the junction triggers a modification of
the damping $\alpha$ that is linear in voltage. For typical Fermi
energy $E_{F}^{L/R}$ of the order of several electron-volts, the
voltage dependence of $\alpha$ may become very notable. This voltage
driven effect may be expressed in terms of $\alpha_0$ and
$\alpha_1(V)$ with $ \alpha(V)=\alpha_0 + \alpha_1$
(Eq.(\ref{voltage+})). Here \begin{eqnarray}
\alpha_0= \pi \bigl(
J_{LL}^2[D(E_F^L)]^2 + 2J_{LR}^2 D(E_F^L)
D(E_F^R)+J_{RR}^2[D(E_F^R)]^2 \bigl), \nonumber
\\
\alpha_1= \frac{1}{4 e|T_{0}|^{2}[D(E_F^L) + D(E_F^R)]} \Bigl(
\frac{I_{o}^{L}}{[D(E_F^L)]^2} + \frac{I_{o}^{R}}{[D(E_F^R)]^2}
\Bigl) \nonumber
\\ \times \bigl[ D(E_F^L)
D(E_F^R) \bigl( J_{LL}^{2} \frac{\partial D^{L}(E_F)}{\partial
E_{F}} - J_{RR}^{2} \frac{\partial D^{R}(E_F)}{\partial E_{F}}
\bigl) \nonumber
\\+ J_{LR}^{2} \bigl( [D(E_F^R)]^2 \frac{\partial D^{L}(E_F)}{\partial E_{F}}
- [D(E_F^L)]^2 \frac{\partial D^{R}(E_F)}{\partial E_{F}} \bigl)
\bigl].
\label{central++}
\end{eqnarray}


\section{Conclusions}
In conclusion, we present a theoretical study of
Landau-Lifshitz-Gilbert damping (Eq.(\ref{LLG2})) for a localized
spin ${\vec S}$ in a junction. The exchange interactions between the
localized spin and tunneling electrons leads to additional
dissipation of the spin motion, see Fig.(\ref{FIG:FIG2}). In the
presence of an applied voltage bias $V$, the damping coefficient,
i.e., Gilbert damping, is modified in linear order in $V$ for the
leads with particle-hole asymmetry in the Density of States.

\section{Acknowledgements}

Work at LANL was supported by the US DOE under LDRD X1WX.

\section{Appendix A: DERIVATION OF THE EFFECTIVE ACTION}
Here we will give a detailed derivation of the effective action for
a spin. From Eq.(\ref {eff}), we can extract a quadratic form of
spins with the aid of the well known identity ${\rm {ln}}~{\rm
{det}}M ={\rm Tr}~{\rm {ln}}M.$  In order to tabulate the expansion
of ${\rm Tr}~{\rm {ln}}M$ perturbatively, we define matrices $M_0$
and $M_1$,

\begin{eqnarray}
M_{KP}&&=(M_0)_{KP}+(M_1)_{KP}
\nonumber
\\
(M_0)_{KP} &&\equiv ((-i\omega + \xi_{\alpha k})\delta_{\alpha \beta} \delta_{\sigma \sigma'})\delta_{KP}
\nonumber
\\
(M_1)_{KP} &&\equiv {\frac{1}{\sqrt{\beta}}}(T_{\alpha \beta}(\omega-\omega'))_{\sigma \sigma'},
\nonumber
\end{eqnarray}
where $K \equiv (k, \omega)$ and $P \equiv (p, \omega')$ with fermionic Matsubara frequencies $\omega$ and $\omega'$.
Employing the expansion ${\rm ln}(1+x) = - \sum_{n=1}^{\infty}
{\frac {(-x)^n}{n}}$, we can  write the effective action as
\begin{equation}
S_{\rm {eff}}(\vec S(\tau)) \sim S_0 + {\rm {Tr}}~{\rm {ln}}M_0 +
{\rm {Tr}}(M_0^{-1}M_1) -{\frac{1}{2}}{\rm {Tr}} (M_0^{-1}M_1)^2 +
\cdots, \label{Seff}
\end{equation}
where $S_0$ is the sum of the first and the second term in
Eq.(\ref{eff}). The third term in Eq.(\ref{eff}) (and consequent
last term shown in Eq.(\ref{Seff})) is the first non-trivial
contribution to the spin equation of motion.
Its evaluation is straightforward,
\begin{eqnarray}
&&{\rm Tr}(M_0^{-1}M_1)^2
\nonumber
\\
&&= \sum (M_0^{-1})_{K_1 K_1}(M_1)_{K_1 K_2}(M_0^{-1})_{K_2 K_2}(M_1)_{K_2 K_1}
\nonumber
\\
&&={\frac{1}{2 \beta}}\sum (-i\omega + \xi_{\alpha k})^{-1} (T_{\alpha \beta}(\omega-\omega'))_{\sigma \sigma'}
\nonumber
\\
&&(-i\omega' + \xi_{\beta p})^{-1} (T_{\beta \alpha}(\omega'-\omega))_{\sigma' \sigma},
\nonumber
\end{eqnarray}
where, repeated indices are implicitly summed over.
Then, we find
\begin{eqnarray}
&&\Delta S = {\frac{1}{2}} {\rm Tr}(M_0^{-1}M_1)^2
\nonumber
\\
&&={\frac{1}{2 \beta}} \sum_{\omega\omega'} \sum_{\alpha \beta} \sum_{\sigma \sigma'}
\sum_{k \in \alpha} \sum_{p \in \beta}(-i\omega + \xi_{\alpha k})^{-1} (T_{\alpha \beta}(\omega-\omega'))_{\sigma \sigma'}
\nonumber \\
& &(-i\omega' + \xi_{\beta p})^{-1} (T_{\beta \alpha}(\omega'-\omega))_{\sigma' \sigma}
\nonumber
\\
&&=-\sum_{\alpha \beta} \sum_{k \in \alpha} \sum_{p \in \beta} \sum_{\omega_m}
J_{\alpha \beta} J_{\beta \alpha} {\frac{f(\xi_k)-f(\xi_p)}{i\omega_m +\xi_k -\xi_p}}
{\vec S}(-\omega_m) \cdot {\vec S}(\omega_m)
\nonumber
\\
&&\equiv -\int d\tau \int d\tau' {\vec S}(\tau) \cdot {\vec S}(\tau') 2K(\tau-\tau').
\nonumber \end{eqnarray}
Here, $K(\tau-\tau')$ denotes the integral kernel defined in
Eq.(\ref{Kt}). Upon invoking the identity $2{\vec S(\tau)}\cdot{\vec
S({\tau}')} = (({\vec S(\tau)})^2+({\vec S({\tau}')})^2) -({\vec
S(\tau)}-{\vec S({\tau}')})^2$, the effective action becomes that of
Eq.(\ref{locdis}).

\section{APPENDIX B: THE DERIVATION OF THE SPECTRAL DENSITY}
\begin{figure}
\centerline{\includegraphics[width=0.9\columnwidth]{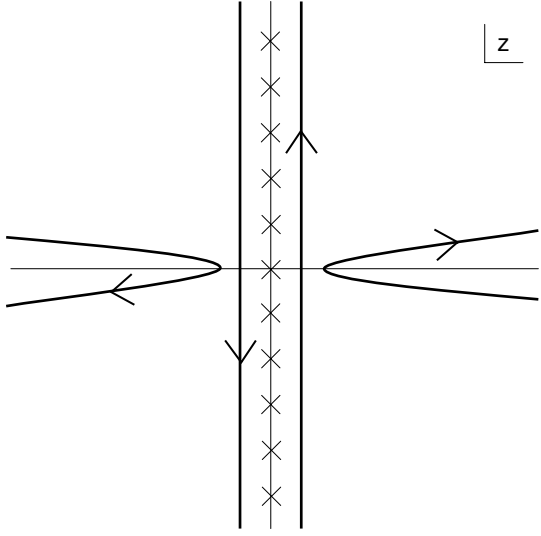}}
\caption{The standard contour employed in Eq.(\ref{cont}) in order
to evaluate the Matsubara sum of Eq.(\ref{Kt}). The crosses along
the imaginary axis denote the Matsubara frequencies.}
\label{FIG:FIG3}
\end{figure}

We return to Eq.(\ref{Kt}) derived in Appendix A, and express the
sum as a contour integral following standard procedures,
e.g.\cite{Mahan}, to obtain
\begin{widetext}
\begin{eqnarray}
K_{\alpha \beta} &=& {\frac{J_{\alpha \beta} J_{\beta \alpha}}{2}}
\sum_{k \in \alpha}  \sum_{p \in \beta} \oint {\frac{dz}{2 \pi i}}
\Bigl( {\frac {e^{-z \tau}}{e^{\beta z}-1}} \theta(-\tau) + {\frac
{e^{-z \tau}}{1-e^{-\beta z}}} \theta(\tau) \Bigr) {\frac
{f(\xi_k)-f(\xi_p)}{z +\xi_k -\xi_p}} \nonumber
\\
&=& P \int_{- \infty}^{\infty} d{\omega} \Bigl( {\frac {J_{\alpha
\beta} J_{\beta \alpha}}{2}} \Bigr)
              \sum_{k \in \alpha}  \sum_{p \in \beta} [f(\xi_k)-f(\xi_p)]
\delta(\omega + \xi_k - \xi_p)
              \Bigl({\frac {e^{-\omega \tau}}{e^{\beta \omega}-1}}
\theta(-\tau) + {\frac {e^{-\omega \tau}}{1-e^{-\beta \omega}}}
\theta(\tau)  \Bigr) \nonumber
\\
&=& P \int_0^{\infty} d{\omega} \Bigl( -{\frac {J_{\alpha \beta}
J_{\beta \alpha}}{2}} \sum_{k \in \alpha}  \sum_{p \in
\beta}[f(\xi_k)-f(\xi_p)] \delta(\omega + \xi_k - \xi_p)  \Bigr)
\frac{{\rm cosh}({{\beta}/2}-|\tau|){\omega}}{{\rm sinh}({\beta
\omega}/2)} \nonumber
\\
&\equiv&  \int_0^{\infty} d{\omega} J(\omega) \frac{{\rm
cosh}({{\beta}/2}-|\tau|){\omega}}{{\rm sinh}({\beta \omega}/2)}
\label{cont}
\end{eqnarray}
\end{widetext}
where $J(\omega)$ denotes the \textit{spectral density} in
Caldeira-Leggett theory. The standard contour employed here is shown
in Fig.(\ref{FIG:FIG3}). The symbol $P$ in Eq.(\ref{cont}) denotes
the principal part of the integral.

The spectral density $J(\omega)$ in Eq.(\ref{spec2})
\begin{widetext}
\begin{eqnarray}
J(\omega)=\frac{J_{LL}^2}{2} \int_{-E_F^L}^{\infty} d{\xi_L}
N(\xi_L) \int_{-E_F^L}^{\infty} d{\xi_L'} N(\xi_L')
[f(\xi_L)-f(\xi_L')]\delta(\omega+{\xi_L}-{\xi_L'}) \nonumber
\\
+\frac{J_{RR}^2}{2} \int_{-E_F^R}^{\infty} d{\xi_R} N(\xi_R)
\int_{-E_F^R}^{\infty} d{\xi_R'} N(\xi_R')
[f(\xi_R)-f(\xi_R')]\delta(\omega+{\xi_R}-{\xi_R'}) \nonumber
\\
+J_{LR}^2 \int_{-E_F^L}^{\infty} d{\xi_L} N(\xi_L)
\int_{-E_F^R}^{\infty} d{\xi_R} N(\xi_R)
[f(\xi_L)-f(\xi_R)]\delta(\omega+{\xi_L}-{\xi_R}) \nonumber
\\
\sim  \Bigl( \frac{J_{LL}^2}{2} N(\xi_L =0) N(\xi_L' =0) +
\frac{J_{RR}^2}{2} N(\xi_R =0) N(\xi_R' =0)
+ J_{LR}^2 N(\xi_L =0) N(\xi_R =0) \Bigr) \omega
\nonumber
\\
=\Bigl( \frac{J_{LL}^2}{2} D^L(E_F^L) D^L(E_F^L) + \frac{J_{RR}^2}{2}
D^R(E_F^R) D^R(E_F^R)
            + J_{LR}^2 D^L(E_F^L) D^R(E_F^R) \Bigr) \omega,
\label{aspec}
\end{eqnarray}
\end{widetext}

where $D^{L/R}(E^{L/R}_F)$ denotes the \textit{density of states} at the Fermi
energy level of the left/right lead.
\\
If we apply a voltage leading to a chemical potential drop of
$E^L_F-E^R_F=(E_F +\Delta\mu_L)-(E_F+\Delta\mu_R)=eV$, then
Eq.(\ref{spec}) follows. This, in turn, leads to Eq.(\ref{Gil}).
then Eq.(\ref {spec}) follows.

\section{APPENDIX C: THE SPIN EQUATION OF MOTION}

If $J(\omega) = \frac{\alpha \omega}{2 \pi}$ then, from
Eq.(\ref{K}), the non-local in time kernel of the action
(Eq.(\ref{K+})) is $K(\tau) \sim {\frac{\alpha}{2 \pi}}
{\frac{1}{{\tau}^2}}$. We  thus obtain from Eq.(\ref{locdis}),
\begin{equation}
\Delta S_{dis} =  {\frac{\alpha}{2 \pi}} \int d{\tau} \int d{\tau}'
{\frac {({\vec S(\tau)}-{\vec S({\tau}')})^2} {(\tau-{\tau}')^2}}.
\end{equation}
The functional derivative of $\Delta S_{\rm {dis}}$ with respect to
$\vec S(\tau)$ is
\begin{eqnarray}
{\frac{\delta \Delta S_{dis}}{\delta S(\tau)}} =
{\frac{\alpha}{\pi}} \int d{\tau}'{\frac {({\vec S(\tau)}-{\vec
S({\tau}')})} {(\tau-{\tau}')^2}} \nonumber
\\ = {\frac{\alpha}{\pi}} \int d{\tau}' {\frac{1}{(\tau-{\tau}')}}
{\frac{d}{d{\tau}'}}({\vec S(\tau)}-{\vec S({\tau}')}) = i \alpha
{\frac {d}{d{\tau}}}{\vec S(\tau)} \label{corr}
\end{eqnarray}
From the free portion of the action (the first two terms of
Eq.(\ref{eff})), we have
\begin{equation}
{\frac{\delta S_0}{\delta {\vec S(\tau)}}} = i {\frac{1}{S^2}}
{\frac{d{\vec S(\tau)}}{d{\tau}}} \times {\vec S(\tau)} + \vec h.
\label{free}
\end{equation}
Adding Eqs.(\ref{corr},\ref{free}), equating the sum to zero, cross
multiplying with $\vec{S}(\tau)$, and changing $\tau \to it$, we
obtain Eq.(\ref{LLG2}).

\end{document}